\documentclass[reprint,floatfix, superscriptaddress, preprintnumbers, aip]{revtex4-2}
\usepackage[]{graphicx}
\usepackage{epstopdf}
\usepackage{array}
\usepackage{amsmath}
\usepackage{amssymb}
\usepackage{bm}
\usepackage{hyperref}
\usepackage{framed}
\usepackage{MnSymbol}
\usepackage{float}
\usepackage[dvipsnames]{xcolor}
\usepackage{ulem}
\usepackage{dcolumn}
\usepackage{bm}
\usepackage{orcidlink}
\date{\today}

\begin{document}
\title{Development of a high-field pulsed magnet and optical fiber coupled cryostat system for magneto-photoluminescence measurements}
\author{Deepesh Kalauni\,\orcidlink{0009-0007-4930-4709}}
\affiliation{Indian Institute of Science Education and Research Kolkata, Mohanpur, Nadia 741246, West Bengal, India}

\author{Kingshuk Mukhuti\,\orcidlink{0000-0003-1333-6307} }
\affiliation{Indian Institute of Science Education and Research Kolkata, Mohanpur, Nadia 741246, West Bengal, India}

\author{Tao Peng\,\orcidlink{0000-0002-8287-3723}}
\affiliation{School of Electrical and Electronic Engineering, Wuhan National High Magnetic Field Center, Huazhong University of Science and Technology, Luoyu Road 1037, Wuhan, China}


\author{Bhavtosh Bansal\,\orcidlink{0000-0002-7003-5693}}\email{bhavtosh@iiserkol.ac.in}
\thanks{Corresponding author}
\affiliation{Indian Institute of Science Education and Research Kolkata, Mohanpur, Nadia 741246, West Bengal, India}

\date{\today}
\begin{abstract}
This paper presents the development of a high field pulsed magnet system combined with an optical fiber-coupled low temperature cryostat for magneto-photoluminescence measurements. A novel aspect of our system is the use of electrolyte capacitors in a 75 kJ capacitor bank to drive the pulsed magnet, enabling the generation of high magnetic fields at a relatively low charging voltage (400 V). The zylon reinforced wire wound magnet coil achieves a maximum field strength of 35 tesla with a magnetic field rise time of 10 ms. The integrated cryostat was developed to meet the specific requirements of magneto-photoluminescence experiments using a 4 K helium closed-cycle cryocooler to provide a stable low-temperature environment down to 5 K. 

\end{abstract}
\maketitle

\section{INTRODUCTION}
High magnetic fields combined with photoluminescence spectroscopy have emerged as essential tools for probing the fundamental physics of condensed matter systems, particularly in the investigation of excitonic phenomena and quantum states in low dimensional semiconductors.\cite{crooker2020gan,gen2020crystalfield,qiang2021polarized,shornikova2025brightdark,zhang2023dark,shornikova2025bright,serati2024probing,mintairov2025dirac,shornikova2020magneto, Hayne2012_C2} Strong magnetic fields allow access to energy levels that are otherwise hidden, providing detailed insight into spin and valley physics, many-body interactions, and quantum correlations. Recent years have witnessed remarkable advances in understanding these phenomena through magneto-photoluminescence measurements, with studies demonstrating Landau quantization of excitons in monolayer transition metal dichalcogenides,\cite{liu2020landau} the brightening of spin-forbidden dark excitons in MoS$_2$ and MoSe$_2$ monolayers,\cite{robert2020measurement} and the observation of complex many-body states such as six-body and eight-body exciton complexes in WS$_2$.\cite{vantuan2022sixbody} These studies deepen our understanding of quantum phenomena and facilitate spintronic and valleytronic applications.

Low temperature photoluminescence measurements are essential for resolving excitonic fine structure and quantum phenomena, as suppressing thermal broadening enables the separation of closely spaced energy levels. Moreover, reduced phonon scattering at low temperatures enhances the photoluminescence signal by minimizing non radiative recombination.\cite{Cong2015_Superfluorescence,Bryja2004_MagneticExcitons,Li2019_DarkExciton}

Conventional approaches to generating magnetic fields beyond $\gtrsim$ 15 tesla, including superconducting and resistive DC magnets face significant limitations in accessibility, cost, and operational complexity.\cite{zhang202426,twin2007present,Hollis,chen2012resistive} Indeed while high temperature superconducting solenoids have been demonstrated to reach beyond 26 tesla, typical laboratory superconducting magnets have been restricted to about 12--14 tesla, beyond which the cost becomes prohibitive. 

Pulsed magnets have long offered an effective alternative, achieving non-destructive fields exceeding 100 T.\cite{zherlitsyn2010design} While the magnetic fields are transient, with pulse durations typically in the millisecond to second range, this timescale is well-suited for investigating condensed matter phenomena that occur on picosecond to microsecond timescales.

High magnetic field user facilities worldwide have established magneto-photoluminescence measurements based on pulsed magnetic fields, however these systems typically rely on liquid helium cryostats, which require substantial resources and operational costs.\cite{han2017pulsed,boebinger2001national,tay2022magneto,noe2013tabletop,lncmp2006toulouse} Closed-cycle refrigerators (CCR) offer a practical alternative by achieving temperatures down to 5 K without liquid cryogen consumption, with successful implementations demonstrated across diverse measurement techniques including magnetotransport, Mössbauer spectroscopy, NMR, and SQUID microscopy.\cite{murthy2007construction,naumov2010closed,berryhill2008novel,goto2011optical,low2021scanning}

In this paper, we report complete development of a high pulsed magnetic field system capable of reaching 35 T.\cite{Mukhuti-Bansal} The system  utilizes an electrolytic capacitor bank using a very low charging voltage of 400 V. This is in contrast to conventional pulsed magnets, which typically require voltages in the 3–12 kV range. Furthermore, the system is integrated with a cryogen-free closed-cycle refrigerator (CCR) designed to achieve temperatures down to 5 K at the sample stage, eliminating the need for liquid helium based cooling. To facilitate photoluminescence measurements within the narrow magnet bore, we have implemented a fiber optic coupling approach. This solution eliminates the need for free-space optics inside the magnet, significantly reduces alignment complexity, and prevents beam deflection caused by intense magnetic forces. Collectively, these advancements offer a robust platform for investigating the optical behavior of materials in high magnetic fields and cryogenic environments. Such facilities are usually available only at large national laboratories equipped with substantial resources and personnel.\cite{Kratz2002_C2,Herrmannsdoerfer2003_C2,Singleton2004_C2,Ortenberg2001_C2, Miura2001_C2,Bykov2001_C2} The work demonstrates that a sophisticated system within the constraints of a small laboratory space and budget can also be developed in small laboratories, albeit with a lower peak field. 
\section{The pulsed magnet}
\subsubsection{The generation circuit}
The pulsed magnet is operated using modest a capacitive discharge circuit, shown in Fig. 1. 
\begin{figure}[tb]
	\includegraphics[scale=0.2]{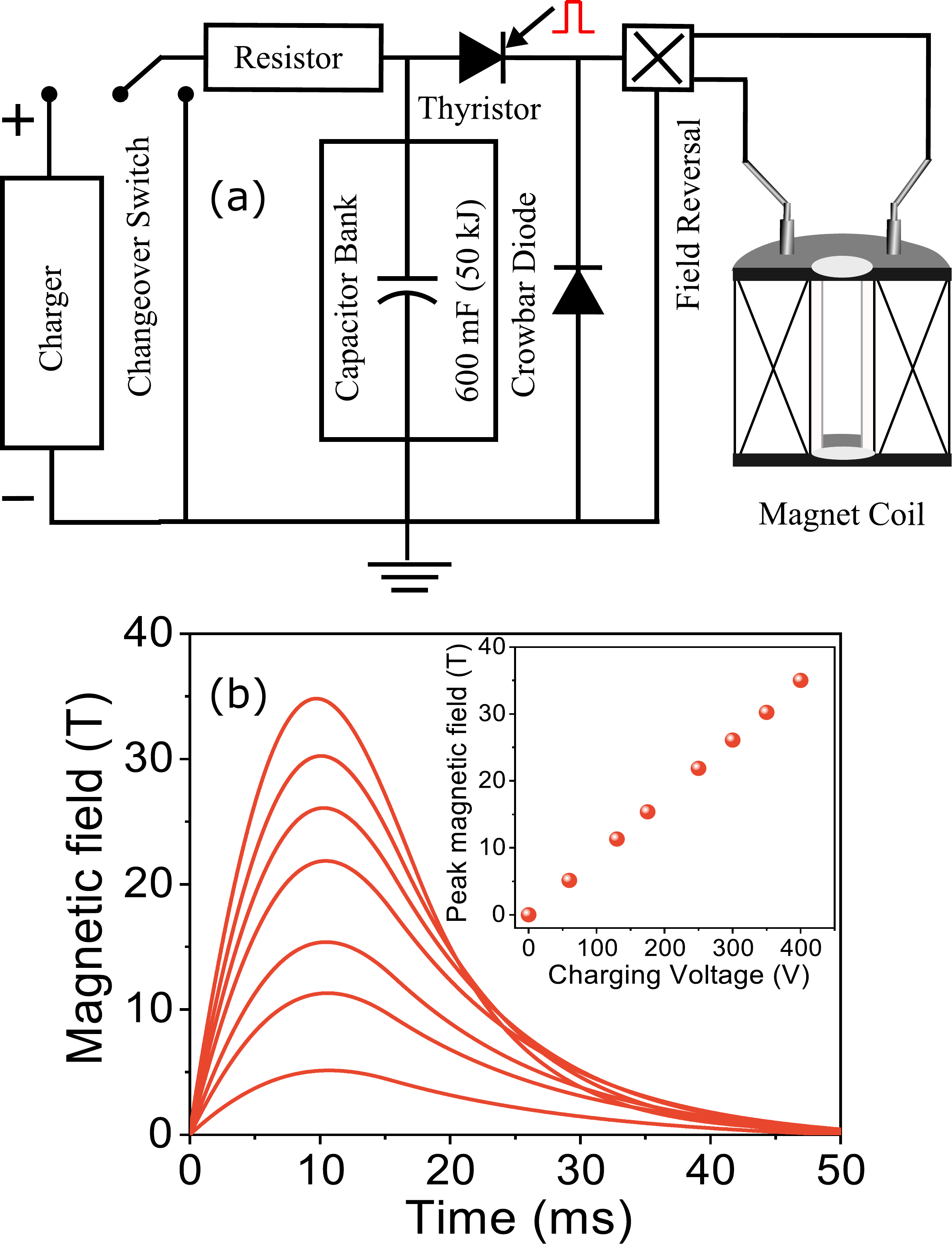}
	\centering
	\caption{(a) Schematic of the pulsed magnetic field generation circuit. (b) Magnetic field profiles recorded at various charging voltages. The inset in panel (b) shows the dependence of the peak magnetic field on the applied charging voltage.}
	\label{Fig1}
\end{figure}
The capacitor bank comprises of 60 standard electrolytic capacitors \textit{(Alcon Electronics Pvt. Ltd.}, India) connected in parallel, each with a capacitance of 10 mF, resulting in a total capacitance of 600 mF. The bank can be charged to a maximum of 500 V. During operation, the capacitors are first charged using a locally made high current unregulated dc power supply, connected to the bank through a high power 75 $\Omega$ current limiting resistor. A changeover switch enables the same resistor to connect the capacitor bank to ground and to gently discharge the capacitors without firing the magnet, if needed. 
Once the bank is charged,  the stored energy is rapidly discharged through the magnet coil by remotely triggering a thyristor, producing a transient magnetic field [Fig. 1(b)]. A crowbar diode prevents the current from oscillating and protects the electrolytic capacitors from damage by reverse charging. A high current double-pole double-throw switch is used for field reversal. The thyristor trigger (a current amplified 5 V TTL pulse connected to the thyristor via a 1:1 isolation transformer) is also used to synchronize the rest of the experiment. 

\begin{table}[!b]
\caption{\label{tab:coil_specs}Specifications of the magnet coil}
\begin{ruledtabular}
\begin{tabular}{lc}
Parameter & Value \\
\hline
Inner bore size & 18 mm \\
Axial length & 60 mm \\
Coil conductor material & Copper, 8-layer winding \\
Wire length & 13.6 m \\
Mechanical reinforcement & Zylon, 7 layers \\
Inductance & 0.138 mH \\
Resistance at 300 K & 23.71 m$\Omega$ \\
Resistance at 77 K & 2.89 m$\Omega$ \\
Maximum magnetic field & 35 T \\
Stored energy at peak field & 48 kJ \\
\end{tabular}
\end{ruledtabular}
\end{table}
\begin{figure*}[!tb]
	\includegraphics[scale=0.35]{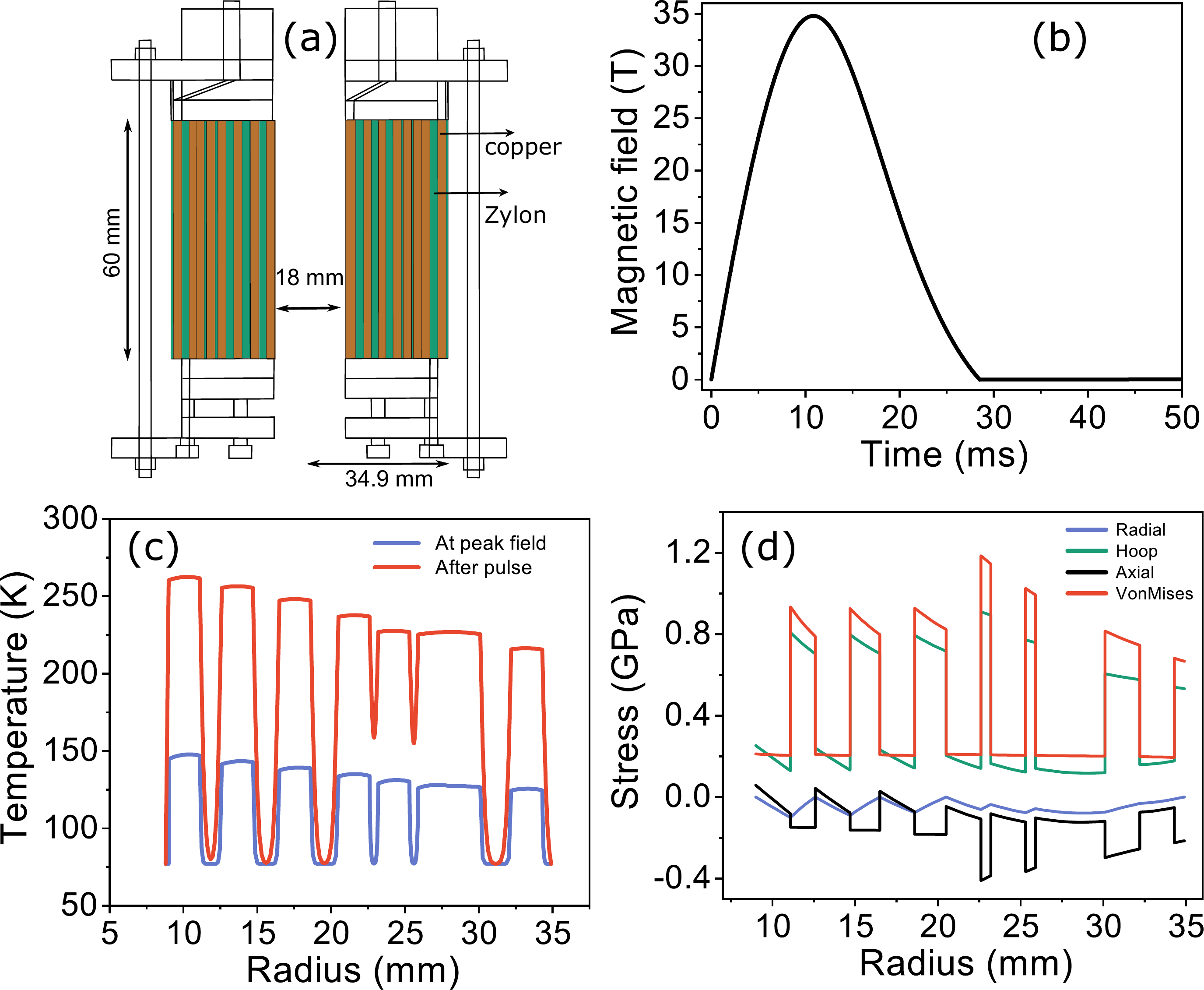}
	\centering
	\caption{Results simulated using the Pulsed Magnet Design Software (PMDS) at a peak magnetic field of 35 T. (a) Schematic of the designed magnet coil, (b) magnetic field profile, (c) radial heating distribution of the coil, and (d) stress distribution as a function of radial distance.}
	\label{Fig1}
\end{figure*}

\subsubsection{The magnet coil}
Fig. 2(a)  shows the schematic of the magnet coil. The coil is designed with a free inner bore of 18 mm to maximize the peak magnetic field under the constraints set the capacitor bank. With simulations performed using the {\it Pulsed Magnet Design Software} (PMDS),\cite{peng2008efficient,herlach2006experimental,vanacken2001pulsed} 35\,T was seen to be possible. Final design has a total of 15 winding layers, including 8 conductive layers and additional zylon reinforcement layers. The maximum achievable field in such a coil is limited by the ultimate tensile strength (UTS) of the materials used, as the conductor must withstand the substantial Lorentz forces produced by high currents and intense magnetic fields. In this design, copper is used as the primary conducting material however, copper alone restricts the peak field to approximately 20\,T due to its mechanical limitations. To overcome this, the conductor is reinforced with a high strength composite fiber. In our coil, zylon is used as the reinforcing material, allowing the system to safely reach higher magnetic fields by significantly enhancing mechanical strength. Fig. 1(b) presents the simulated magnetic field profile, indicating a peak magnetic field of 35 T achieved. Fig. 1(c) presents the simulated temperature distribution in the magnet coil at the peak magnetic field and after the pulse. A significant amount of thermal energy is generated after the decay of the transient current in the coil, as indicated by the red colored region. Fig. 1(d) shows the stress distribution as function of radius in the coil. The specifications of the magnet are summarized in Table I, while photographs of the magnetic coil and capacitor bank are provided in the Supplementary Material. 

Experimental pulse field profiles obtained at various charging voltages are shown in Fig. 1(b). Magnetic field measurements were carried out using a calibrated pickup coil placed at the center of the magnet bore with the induced voltage measurement by the L4534A  (\textit{Agilent Technologies}) digitizer. The digitizer can operate with 4 parallel channels at 16-bit resolution at a maximum sampling rate of 20 MSa/s (simultaneous). A charging voltage of about 400\,V was required to reach a peak magnetic field of 35\,T at the center of the bore. The magnet coil is immersed in liquid nitrogen to reduce overall heating.

\section{Helium cryocooler-based cryostat}

\begin{figure*}[!ht]
	\includegraphics[scale=0.4]{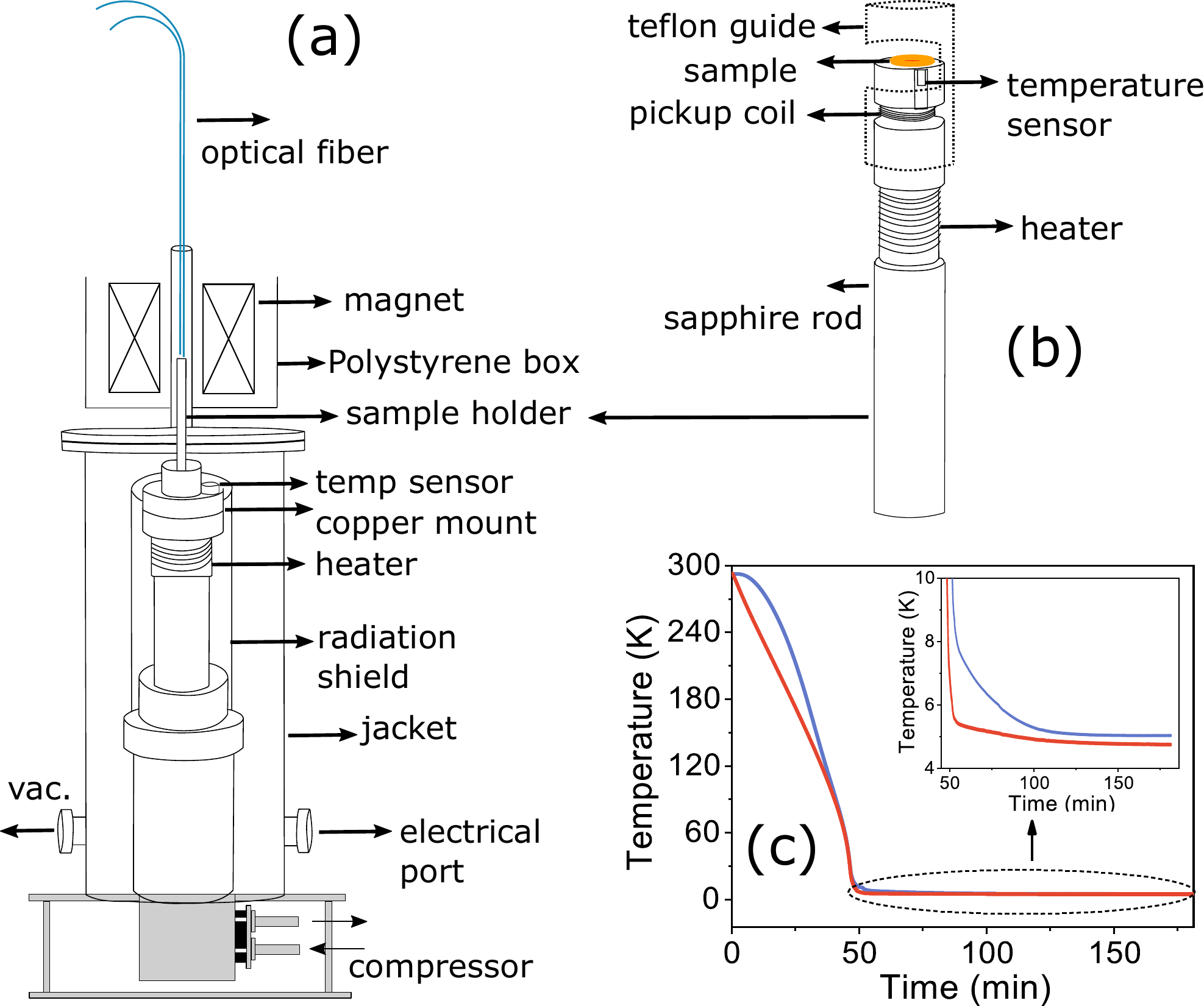}
	\centering
	\caption{(a) Schematic of the low temperature cryostat mounted on the cryocooler cold head, including the sample holder, the temperature sensors, heater coil, aluminum radiation shield, stainless steel vacuum jacket, electrical and vacuum feedthroughs, optical fiber access, and the liquid nitrogen environment within a polystyrene (styrofoam) container. (b) Enlarged schematic view of the sample holder assembly, highlighting the sample, pickup coil, temperature sensors, heater, sapphire rod and teflon guide. (c) The temperature profiles recorded by a sensor placed at the cryocooler cold head and a sensor at the sapphire rod sample mount. The inset provides a magnified view of the low-temperature region. The sapphire rod sensor reaches a base temperature down to 5 K (blue), while the cold head sensor achieves a minimum temperature of 4.6 K (red).}
	\label{Fig1}
\end{figure*}

Fig. 3(a) illustrates the design of our low temperature cryostat, specifically developed to meet the unique requirements of our magneto-photoluminescence experiments. We used a bare GM type closed-cycle helium cryocooler \textit{(Quantum Technology, Canada)} with a relatively large cooling capacity of approximately 1.7 W at 3.5 K, with a specified minimum operating temperature of about 3.8 K at the cold head. A high cooling-capacity cooler was selected due to the relatively large thermal mass of the sample holder and the need to cool the sample located several centimeters away from the physical cold head. A stainless steel shroud,  precisely joined using leak-tight argon tungsten inert gas (TIG) welding was designed and manufactured to be attached to the bare cold head for low-temperature operation. An aluminum radiation shield, made as a single piece for ease of machining and weight reduction, surrounds the cold stage, and it is anchored to the first stage of the cold head, which has a relatively high cooling capacity and cools down to about 50 K. Aluminum is typically chosen because of its low density, relatively high thermal conductivity and strong reflectivity, which help minimize radiative heat load. We note that the performance could have been slightly improved by gold plating. The copper cold head of the cryostat was thermally coupled to a matching copper mount, onto which a sapphire rod was mechanically anchored and coated with thermal grease ({\it Apiezon-N}) to ensure good thermal contact with copper piece.
\begin{figure*}[tb]
	\includegraphics[scale=0.2]{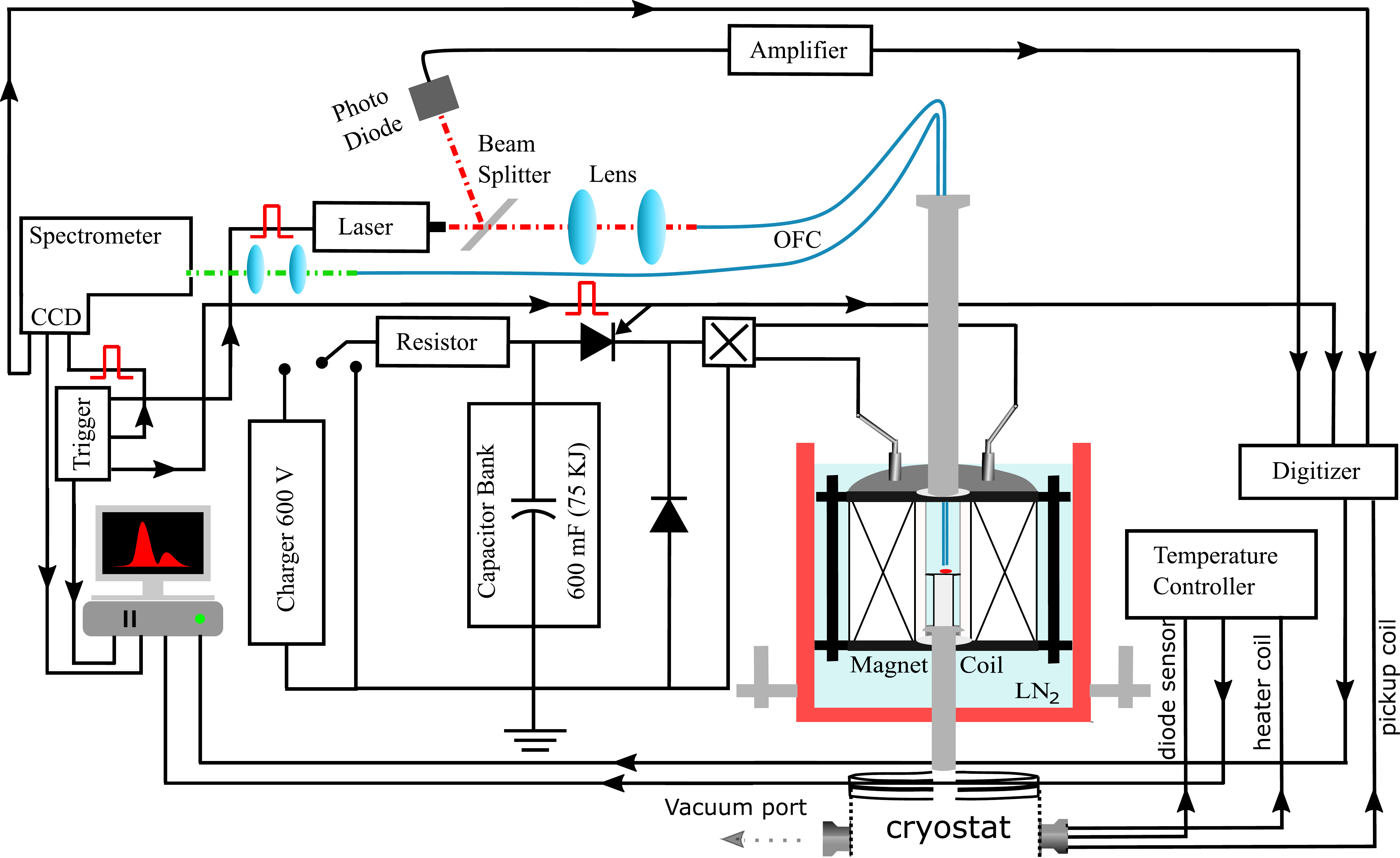}
	\centering
	\caption{Schematic of the complete magneto-PL setup. The magnet coil is immersed in a liquid nitrogen bath, and the cryostat is positioned so that the sample is located at the center of the magnet. A capacitor bank discharges through the coil to generate the magnetic field. Spectroscopic measurements are performed synchronously with the magnetic field pulse. The entire system is operated and monitored remotely.}
	\label{Fig1}
\end{figure*}
Fig. 3(b) presents a detailed view of the sample holder, which is fabricated from sapphire to prevent eddy currents under high pulsed magnetic fields. Sapphire’s electrical insulation combined with excellent thermal conductivity \cite{Richardson-Smith} at low temperature makes it ideal for this application. There is a compromise that one needs to make regarding the length of the sapphire rod. While on the one hand, one would like to maximize the distance of the copper head from the mount, while on the other hand a longer rod would increase the thermal load on the system. Both the length and the load would increase sample temperature, which is undesirable. 
The sapphire rod was chosen to be long enough (12 cm) to ensure that was no electrically conducting part at least within the magnet. All wiring and contacts for the temperature sensors, heater, and pickup coil are carefully integrated into the cryostat. Two temperature sensors ({\it DT670} diode temperature sensor from \textit{Lake Shore Cryotronics}) are employed for precise thermal monitoring: one positioned near the sample to accurately measure the sample temperature, and another mounted on the cold stage to monitor the overall system temperature. For precise temperature control and stable operation, two nichrome wire heaters are used, one located near the cold head neck and the other near the sample in the sapphire sample holder. Optical fibers are coupled into the system to excite the sample and collect the photoluminescence signal. To ensure precise alignment and mechanical stability of the optical fibers, a small teflon guide is placed above the sapphire rod, securely holding the fibers in position. Pictures of the cryostat components and their assembly are provided in the Supplementary Material. Constructing the cryostat was challenging due to several constraints, including the small bore size of the magnet, maintaining vacuum integrity after fiber alignment, and accommodating liquid nitrogen immersion during pulsed field experiments. Despite these difficulties, we have achieved vacuum levels of approximately $10^{-6}$~mbar and temperatures down to 5 K, demonstrating the robustness and effectiveness of the design.

Fig. 3(c)  shows the temperature profile measured by the both cold head sensor and the sapphire rod sensor during natural cooling of the cryostat under vacuum conditions. Both sensors show a gradual decrease in temperature and reach their respective base temperatures within approximately one hour. The cold head sensor reaches a minimum temperature of about 4.2 K, while the temperature at the sample location, measured by the sapphire rod sensor, stabilizes at approximately 5 K. A slight delay in the cooling response of the sapphire rod sensor relative to the cold head sensor is observed, which can be attributed to the additional thermal mass and thermal resistance of the sapphire rod. The cryostat is evacuated using an external pumping system and is able to maintain a stable vacuum level on the order of $10^{-6}\,\mathrm{mbar}$ throughout the measurements, ensuring effective thermal isolation.

\section{Low temperature magnetophotoluminescence}
The integration of a low temperature cryostat with a pulsed magnet system presents several significant technical challenges. The pulsed magnet has a restricted bore diameter of only 18 mm, which severely limits the available space for the sample holder, cryostat jacket, and associated thermal shielding. Maintaining optical access in this cryogenic environment is particularly difficult because the entire magnet is immersed in liquid nitrogen, making free-space optical alignment impractical. Consequently, optical fibers must be used for both excitation and signal collection instead of conventional lens systems, which results in reduced collection efficiency due to the limited numerical aperture and coupling losses in the fibers.

Fig. 4 shows the schematic of the magneto-PL setup. Optical fibers are used for both excitation and collection of the PL signal. This allows the laser, spectrograph, and other optical components to be kept at a safe distance from the magnet. This is particularly important because the fast changing intense magnetic fields can affect sensitive optical equipment and also provide safety in the rare event of a magnet failure or explosion.
In magneto-PL measurements, it is very important to synchronize the laser pulse, magnetic field, and spectrograph exposure, because in a pulsed magnet the magnetic field is present for only a very short time. All three must be synchronized at the peak field, which remains nearly constant for only a small time window (about 1 ms). To achieve this synchronization, we use an inexpensive microcontroller (\textit{Arduino Uno}), which can provide multiple 5 V TTL output pulses. A four channel digitizer is used to verify the timing.
First, both the thyristor and the digitizer are triggered simultaneously. This starts the magnetic field pulse and the digitizer recording. After a set delay, corresponding to the peak field window, TTL pulses are sent to trigger the spectrograph and the laser. Of the remaining digitizer channels, the first records the pickup voltage for measuring the magnetic field profile, the second records the fire output from spectrograph (indicating its acquisition window), and the third records a split portion of the laser signal from a photodiode. The fourth channel is used for the external trigger signal that starts the digitizer.

The photoluminescence (PL) spectroscopy setup uses a diode laser as the excitation source. 
The laser is coupled into a thick multimode optical fiber ($\varnothing$910~\textmu m core, \textit{Thorlabs}) using a microscope objective, which delivers the excitation light to the sample. 
A second optical fiber is used to collect the emitted PL signal from the sample. We used a high power laser diode driven by a controller (ITC4005, \textit{Thorlabs}), which provides stable current and temperature control for reliable operation. The modular laser mount design allows different laser diodes to be easily interchanged, offering flexibility to select excitation wavelengths as required for different measurements.
\begin{figure}[!t]
	\includegraphics[scale=0.18]{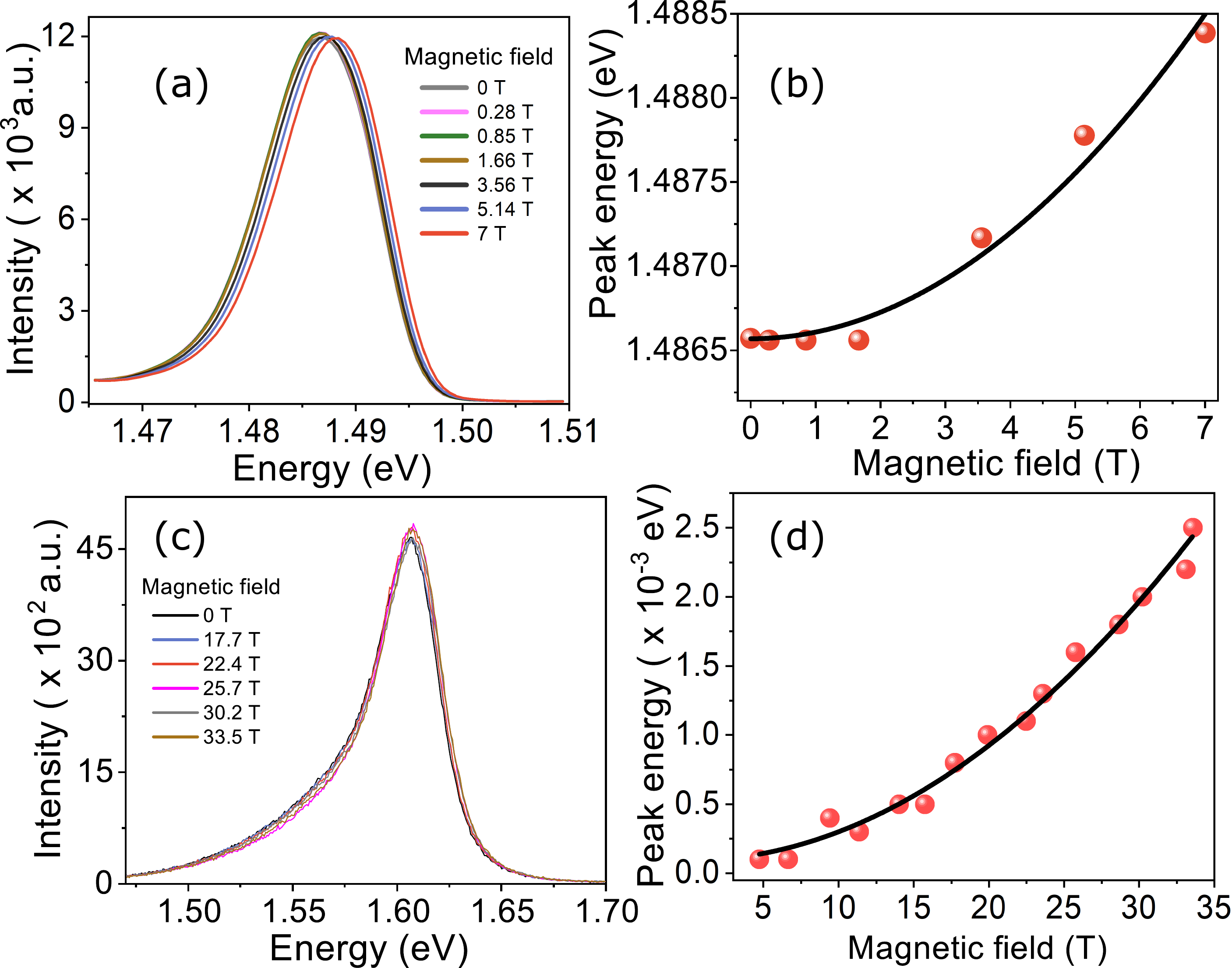}
	\centering
	\caption{Photoluminescence spectra of (a) GaAs and (c) MAPbBr$_3$ single crystals under applied magnetic field. (b,d) Evolution of excitonic peak energy versus magnetic field, fitted with the diamagnetic shift equation (solid lines).}
	\label{Fig1}
\end{figure}
Furthermore,  to accurately determine the magnetic field at the sample position, a pickup coil was placed in close proximity to the sample. This coil was first calibrated using a reference coil of known effective area prepared at the Wuhan laboratory (see Supplementary Material Fig. 3). More direct ways for absolute calibration can also be easily implemented using photoluminescence spectroscopy.\cite{Maes-Diamond}

\subsection{Experimental magnetophotoluminescence data}
To benchmark the system, magneto-photoluminescence measurements were performed at 5 K on two well established excitonic semiconductors used as reference samples, GaAs and MAPbBr$_3$. The diamagnetic shift of the excitonic emission was analyzed to extract the excitonic Bohr radius and binding energy. The measurements were carried out in magnetic fields up to 33.7 T. Fig. 6(a) presents the PL spectra of GaAs under magnetic fields up to 7 T using 638 nm excitation, corresponding to the low magnetic field limit, while the remaining high field data are provided in the Supplementary Material Fig. 5. The emission peak exhibits a clear diamagnetic shift with increasing magnetic field, as illustrated in Fig. 6(b). Fitting the data to the standard quadratic dependence\cite{Klingshirn2012_C5}
\begin{equation}
\Delta E = \frac{q^{2} a_{X}^{2}}{8\mu}\, B^{2}
\end{equation}
yields an excitonic Bohr radius of $a_X \approx 10.9$ nm (using $\mu = 0.067\,m_0$ for GaAs).\cite{moss1961optical,vrehen1968interband}

The corresponding binding energy, calculated from\cite{Patane2012_C2}
\begin{equation}
E_X = \frac{1441.8}{a_X\,\varepsilon_r}
\end{equation}
is $E_X \approx 10.17 $ meV (with $\varepsilon_r = 13$). Both values are in excellent agreement with literature.\cite{Hayne2012_C2}

Fig. 6(c) shows photoluminescence measurements on a high quality MAPbBr$_3$ single crystal under 638~nm excitation. Unlike GaAs, the large exciton binding energy and broad PL line width make the diamagnetic shift barely discernible in the raw spectra \cite{wang2017structural}. However, differential photoluminescence analysis (see Supplementary Material Fig. 4) reveals a systematic shift.

Figure 6(d) presents the extracted diamagnetic shift as a function of magnetic field. The monotonic increase confirms the expected excitonic response. Using the reported values for the reduced mass and dielectric constant ($\varepsilon_r = 16$, $\mu = 0.151\,m_0$),\cite{baranowski2024polaronic, galkowski2016determination} we obtain $a_X \sim 3.8$ nm and $E_X \sim 23.7$ meV, both consistent with previous reports.\cite{tilchin2016hydrogen,yang2017unraveling,PhysRevMaterials.8.034601}

\section{Conclusions}
We have presented the successful design, construction, and integration of a custom optical fiber-coupled cryostat with a homemade pulsed magnet system, enabling magneto-photoluminescence measurements under conditions of high magnetic fields (up to 35 T) and low temperatures (down to 5 K). The system overcomes significant technical challenges inherent to pulsed field experiments, including the restricted 18 mm bore size, suppression of eddy currents through the use of sapphire sample holders, and optical access via fiber coupling. The cryostat achieves vacuum levels of $10^{-6}$ mbar and demonstrates excellent thermal stability. Validation measurements on GaAs and MAPbBr$_3$ samples yielded exciton Bohr radius and binding energy values that are in close agreement with reported literature values, confirming the reliability and precision of our instrumentation. This versatile, cost-effective platform provides a powerful tool for investigating magneto-optical phenomena in semiconductors and opens new avenues for condensed matter research in high magnetic field environments, typically accessible only in large scale facilities.

\section*{ACKNOWLEDGMENTS}
BB thanks V. Venkarataman, Manus Hayne, and Fritz Herlach  for initiating him to the pulsed magnet work. We acknowledge the contributions of many undergraduate students over the years, in particular Sujeet Kumar Choudhary, Md. Arsalan Ashraf, and Mithun S. Prasad. We also thank Pintu Das for workshop support. We thank Uday Kumar and Pradip Khatua for overall help and guidance in design and manufacture of the system.  We thank Siddharth Rout for help with the simulations shown in Fig. 1. This work was partly supported by the ANRF, Department of Science and Technology, Government of India via the grant No. CRG/2018/003282 and No. CRG/2022/008662. DK thanks University Grants Commission, India for PhD fellowship. 

\section*{AUTHOR DECLARATIONS }

\section*{Conflict of Interest }
The authors have no conflicts to disclose.

\section*{Author Contributions}
\textbf{Deepesh Kalauni}: Conceptualization (equal); Data curation (lead); Formal analysis (lead); Investigation (equal); Methodology (equal); Software (lead); Validation (equal); Visualization (equal); Writing -- original draft (lead). \textbf{Kingshuk Mukhuti}: Formal analysis (supporting); Investigation (supporting); Software (lead); Visualization (equal); Writing -- review \& editing (equal). \textbf{Tao Peng}: Methodology (equal); Software (lead); Resources (equal); Writing -- review \& editing (equal). \textbf{Bhavtosh Bansal}: Conceptualization (lead); Funding acquisition (lead); Investigation (lead); Methodology (lead); Project administration (lead); Resources (lead); Supervision (lead); Validation (lead); Formal analysis (lead); Writing -- review \& editing (lead).

\section*{DATA AVAILABILITY}
The data that support the findings of this study are available from the corresponding author upon reasonable request.

\section*{references}


\setcounter{figure}{0}
\setcounter{section}{0}

\vspace{2cm}
\onecolumngrid
\begin{center}
\textbf{\LARGE Supplemental Material}
\end{center}

\section{Images of magnet coil and capacitor bank}
  \begin{figure}[H]
	\includegraphics[scale=0.5]{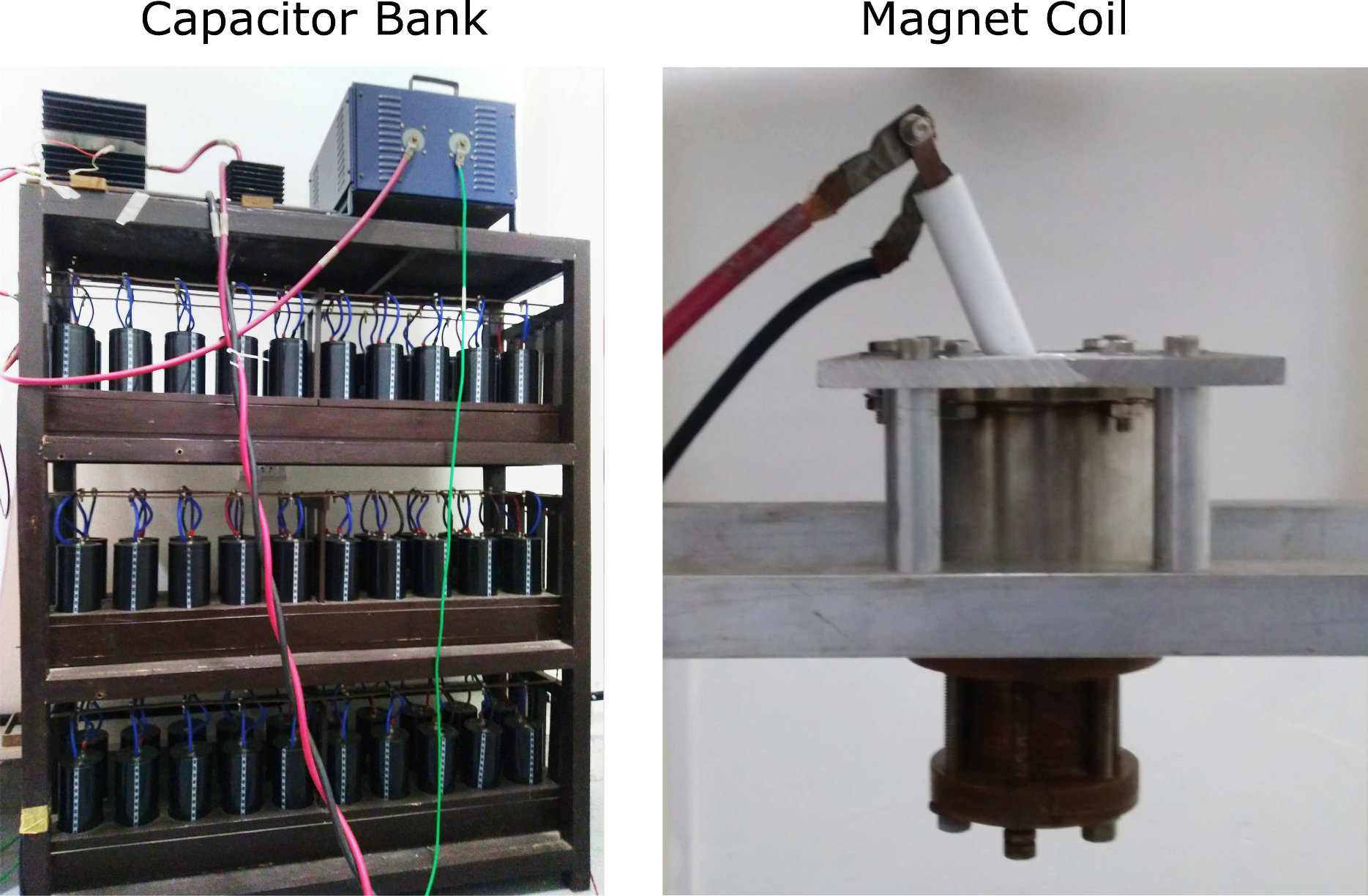}
	\centering
	\caption{Pictures of capacitor bank and magnet coil}
	\label{Fig1}
\end{figure}

 \newpage

 \section{Images of the components of the cryostat and their assembly}

\begin{figure}[H]
	\includegraphics[scale=0.9]{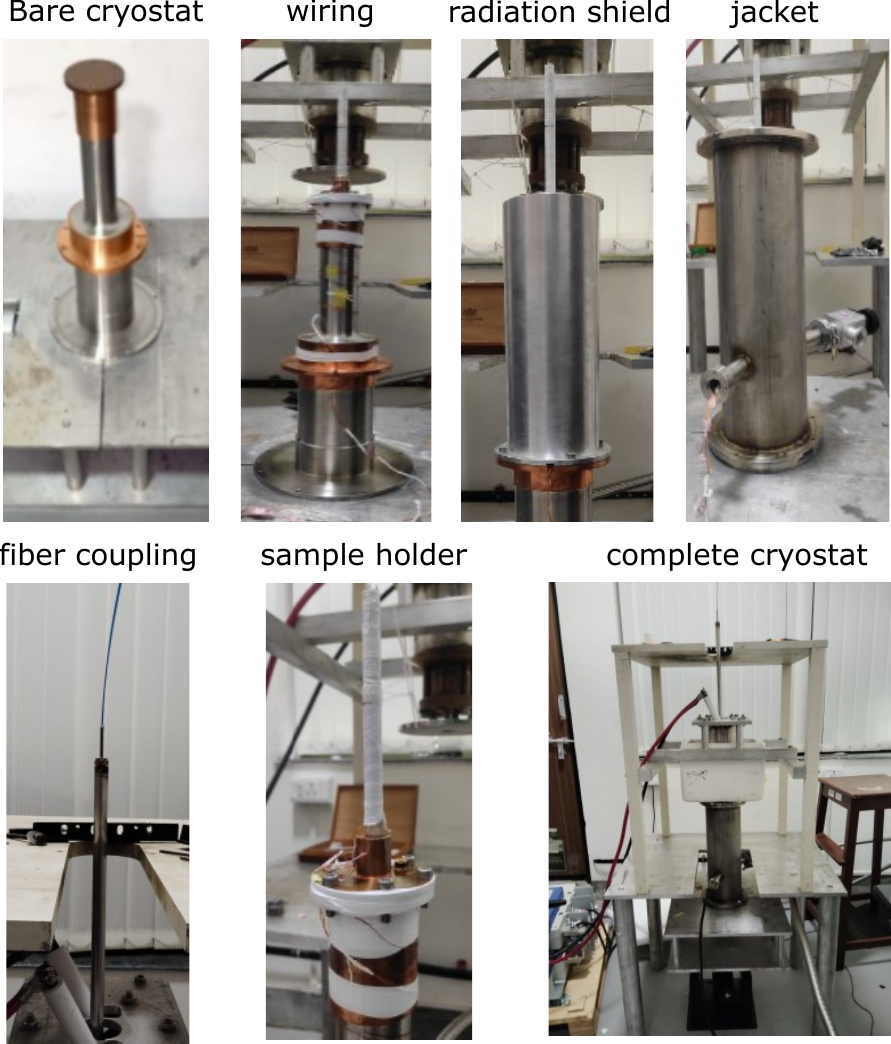}
	\centering
	\caption{Clockwise from top left, the images illustrate the assembly process: the bare cryostat mounted on a mechanical aluminum plate, the wiring connected to the cryostat and sample holder, the cryostat enclosed within the vacuum shroud, the optical fiber coupling to the cryostat, the mounted sample holder, and finally the fully assembled cryostat integrated with the pulsed magnet system}
	\label{Fig1}
\end{figure}

\section{Pickup coil calibration}
\begin{figure}[b]
	\includegraphics[scale=0.3]{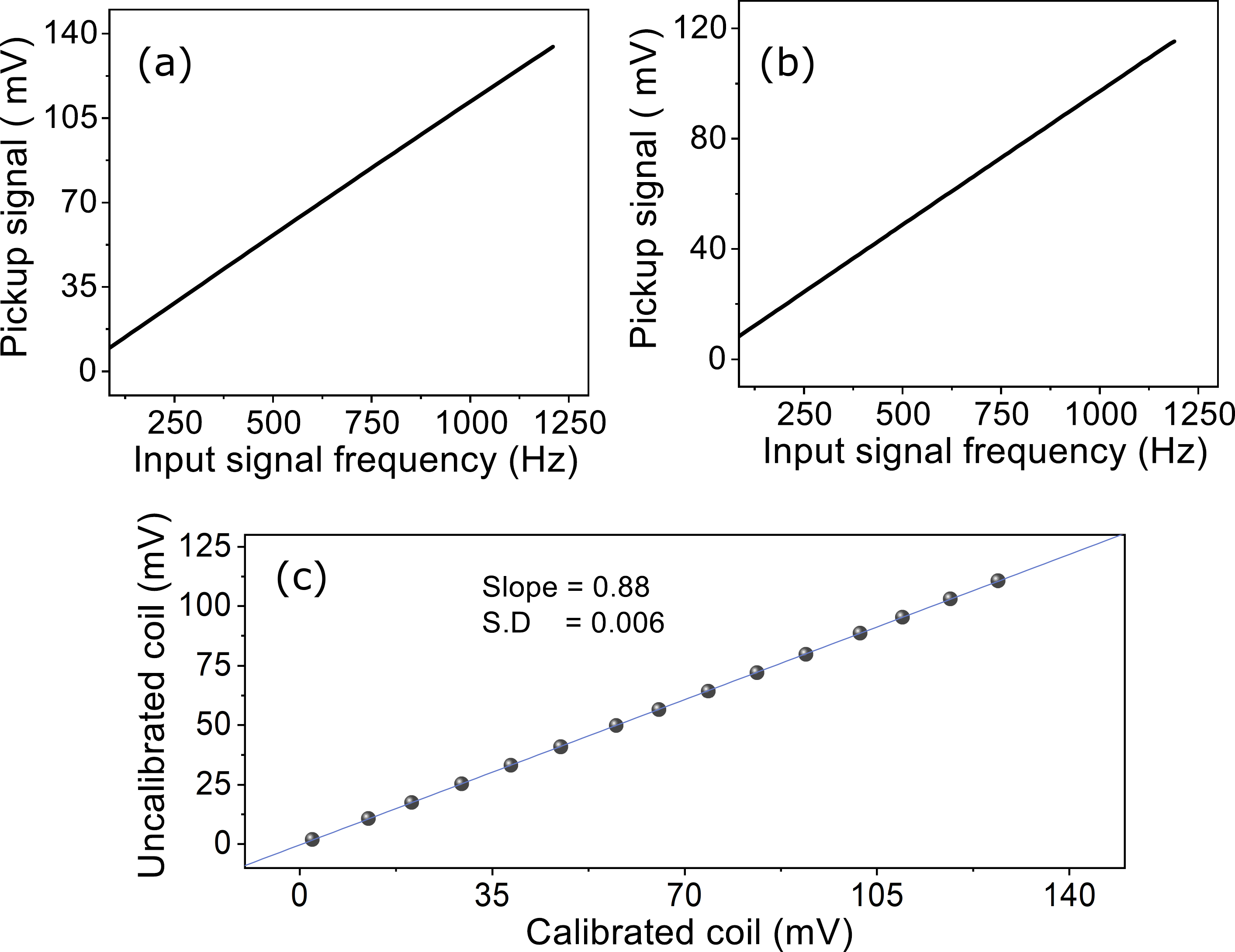}
	\centering
	\caption{(a) Variation of pickup signal with input signal frequency for a calibrated coil at 10 V. (b)  Variation of pickup signal with input signal frequency for a uncalibrated coil at 10 V. (c) Induced voltage ratio between calibrated and uncalibrated pickup coils.}
	\label{Fig1}
\end{figure}

A pre-calibrated coil (provided by Dr. Tao Peng, Wuhan Magnet Lab, China) was used to calibrate the magnetic field. An AC current of varying frequency was applied to the magnet, and the induced voltages in both calibrated and non-calibrated coils were recorded [Figure 1.5(a,b)]. The ratio of their induced voltages gives the ratio of their effective areas. Using this, the area of the non-calibrated coil was calculated [Figure 2.5(c)]. 


 According to Faraday’s law of electromagnetic induction, the voltage induced in a coil due to a time-varying magnetic field is given by:
\begin{equation}
V_i(t) = -A \frac{\partial B(t)}{\partial t}
\end{equation}

where $V_i(t)$ is the induced voltage, $A$ is the area of the coil, and $B(t)$ is the magnetic field as a function of time.

Integrating this expression over time yields the magnetic field:

\begin{equation}
B(t) = \frac{-1}{A} \int_0^t V_i(t) \, dt
\end{equation}

The calibrated area of the pickup coil was found to be $465.32\ \mathrm{mm}^2$.

	\section{Differential photoluminescence analysis for calculating the diamagnetic shift in a MAPbBr$_3$ single crystal}

    \begin{figure}[H]
	\includegraphics[scale=0.4]{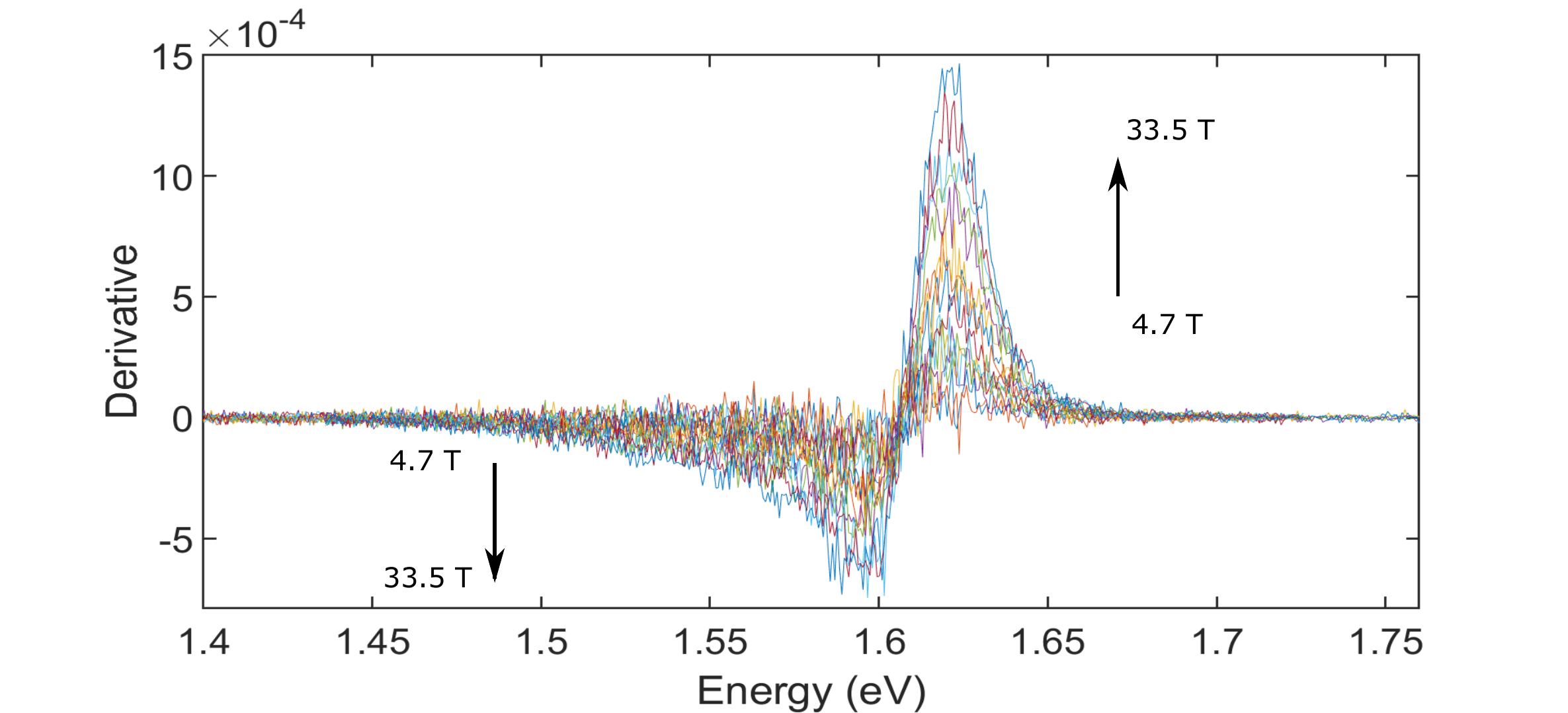}
	\centering
	\caption{Differential PL spectra calculated as the difference between normalized field dependent and zero-field reference spectra for 15 magnetic field strengths ranging from 4.7 T to 33.5 T. The characteristic derivative like lineshape with negative and positive lobes indicates a systematic blue shift of the emission peak due to the excitonic diamagnetic effect. The amplitude of the differential signal increases monotonically with magnetic field strength.}
	\label{Fig1}
\end{figure}


     Photoluminescence spectra were recorded at multiple applied  magnetic field strengths. For each field value, the zero field PL intensity $I_{0}(E)$ was obtained by averaging the five spectra measured at zero magnetic field, while the field dependent spectrum is denoted as $I_{B}(E)$. Because of the large exciton binding energy and the relatively broad PL linewidth, the excitonic diamagnetic shift is barely discernible in the spectra. However, a systematic diamagnetic shift can still be resolved using differential photoluminescence analysis. For each magnetic field strength, the normalized PL spectrum measured in the presence of magnetic field, $I_{B}(E)$, was compared with the normalized zero field reference spectra measured immediately before and after the magnetic field pulse under zero-field ($B = 0$) conditions. Here, \(I_{Z1}(E)\), \(I_{Z2}(E)\), \(I_{Z3}(E)\), \(I_{Z4}(E)\), and \(I_{Z5}(E)\) denote the normalized photoluminescence spectra measured in the absence of a magnetic field.

\[
I_{0}(E)=\frac{I_{Z1}(E)+I_{Z2}(E)+I_{Z3}(E)+I_{Z4}(E)+I_{Z5}(E)}{5}.
\]
The differential spectrum was then calculated as
\[
\Delta I(E)=I_{B}(E)-I_{0}(E)
\]

The differential photoluminescence spectra presented in Fig.~2 provide direct visual evidence of magnetic field induced spectral shifts in MAPbBr$_3$. All 15 differential spectra shows the expected derivative like lineshape, with the zero crossing point remaining approximately constant for all field strengths, confirming that the spectral shift is rigid. The amplitude of the differential signal increases monotonically with increasing magnetic-field strength, reflecting the progressive enhancement of the diamagnetic shift.

\section{High Field Magneto-Photoluminescence Data of Gallium arsenide}

  \begin{figure}[H]
 	\includegraphics[scale=0.4]{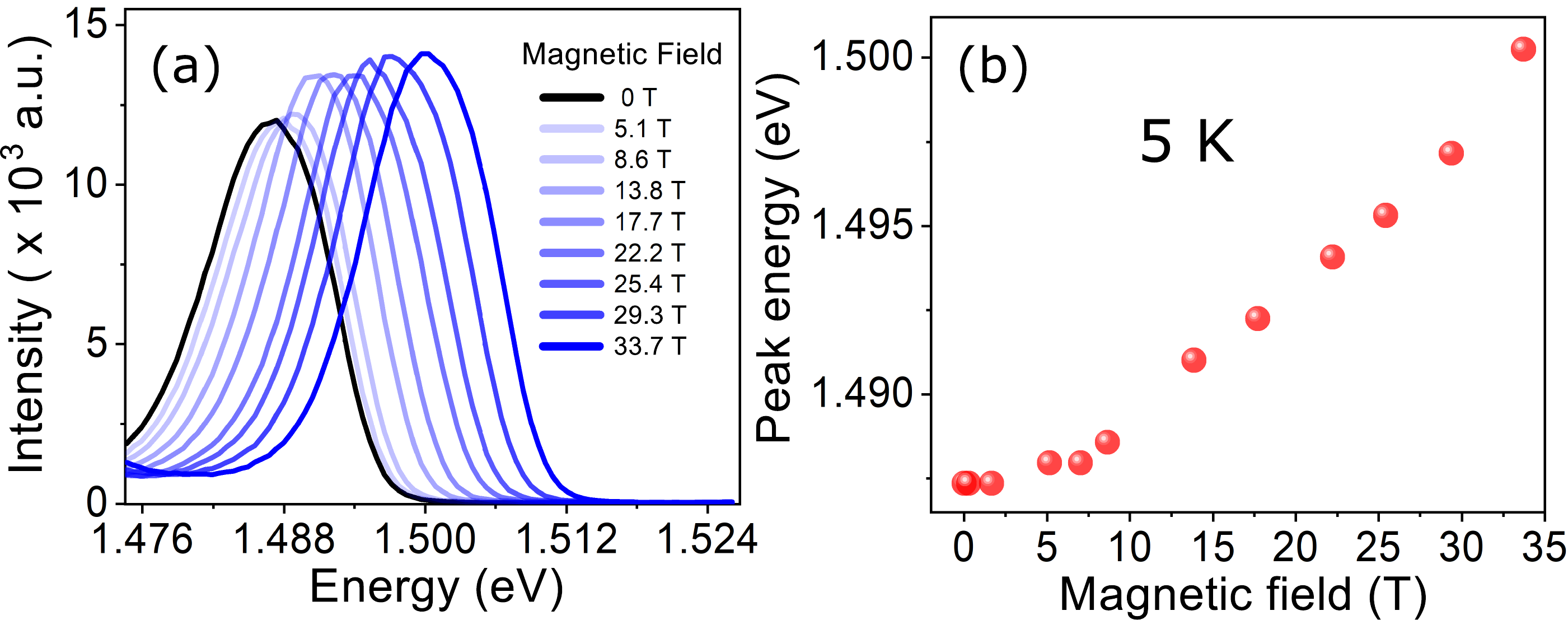}
 	\centering
 	\caption{(a) Photoluminescence spectra of GaAs under applied magnetic fields up to 33.7 T. (b) Evolution of excitonic peak energy versus magnetic field.}
 	\label{Fig1}
 \end{figure}


\begin{thebibliography}{99}

\bibitem{crooker2020gan}
S. A. Crooker, M. Lee, R. D. McDonald, J. L. Doorn, I. Zimmermann, Y. Lai, L. E. Winter, Y. Ren, Y.-J. Cho, B. J. Ramshaw, H. G. Xing, and D. Jena, ``GaN/AlGaN 2DEGs in the quantum regime: Magneto-transport and photoluminescence to 60 tesla,''
\href{https://pubs.aip.org/aip/apl/article/117/26/262105/39065/GaN-AlGaN-2DEGs-in-the-quantum-regime-Magneto}
{\textcolor{blue}{Appl. Phys. Lett.}} \textbf{117}(26), 262105 (2020).

\bibitem{gen2020crystalfield}
M. Gen, T. Kanda, T. Shitaokoshi, Y. Kohama, and T. Nomura, ``Crystal-field Paschen-Back effect on ruby in ultrahigh magnetic fields,''
\href{https://link.aps.org/doi/10.1103/PhysRevResearch.2.033257}
{\textcolor{blue}{Phys. Rev. Res.}} \textbf{2}(3), 033257 (2020).

\bibitem{qiang2021polarized}
G. Qiang, A. A. Golovatenko, E. V. Shornikova, D. R. Yakovlev, A. V. Rodina, E. A. Zhukov, and M. Bayer, ``Polarized emission of CdSe nanocrystals in magnetic field: the role of phonon-assisted recombination of the dark exciton,''
\href{https://pubs.rsc.org/en/content/articlehtml/2020/nr/d0nr07117j}
{\textcolor{blue}{Nanoscale}} \textbf{13}(2), 790–800 (2021).

\bibitem{shornikova2025brightdark}
E. V. Shornikova, D. R. Yakovlev, D. O. Tolmachev, M. A. Prosnikov, P. C. M. Christianen, S. Shendre, F. Isik, S. Delikanli, H. V. Demir, and M. Bayer, ``Bright-dark exciton interplay evidenced by spin polarization in CdSe/CdMnS nanoplatelets for spin-optronics,''
\href{https://pubs.acs.org/doi/full/10.1021/acsanm.4c05364}
{\textcolor{blue}{ACS Appl. Nano Mater.}} \textbf{8}(2), 974–984 (2025). 

\bibitem{Hayne2012_C2}
M. Hayne and B. Bansal, ``High-field magneto-photoluminescence of semiconductor nanostructures,'' 
\href{https://analyticalsciencejournals.onlinelibrary.wiley.com/doi/10.1002/bio.2342}
{\textcolor{blue}{Luminescence}} \textbf{27}(3), 179--196 (2012).
\bibitem{zhang2023dark}
C. Zhang, X. Jiang, P. C. Sercel, H. Lu, M. C. Beard, S. McGill, D. Semenov, and Z. V. Vardeny, ``Dark exciton in 2D hybrid halide perovskite films revealed by magneto-photoluminescence at high magnetic field,''
\href{https://advanced.onlinelibrary.wiley.com/doi/full/10.1002/adom.202300436}
{\textcolor{blue}{Adv. Optical Mater.}} \textbf{11}(18), 2300436 (2023).

\bibitem{shornikova2025bright}
E. V. Shornikova, D. R. Yakovlev, D. O. Tolmachev, M. A. Prosnikov, P. C. M. Christianen, S. Shendre, F. Isik, S. Delikanli, H. V. Demir, and M. Bayer, ``Bright-dark exciton interplay evidenced by spin polarization in CdSe/CdMnS nanoplatelets for spin-optronics,''
\href{https://pubs.acs.org/doi/full/10.1021/acsanm.4c05364}
{\textcolor{blue}{ACS Appl. Nano Mater.}} \textbf{8}(2), 974–984 (2025).

\bibitem{serati2024probing}
C. Serati de Brito, B. L. T. Rosa, A. Chaves, C. Cavalini, C. R. Rabahi, D. F. Franco, M. Nalin, I. D. Barcelos, S. Reitzenstein, and Y. G. Gobato, ``Probing the nature of single-photon emitters in a WSe$_2$ monolayer by magneto-photoluminescence spectroscopy,''
\href{https://pubs.acs.org/doi/full/10.1021/acs.nanolett.4c03686}
{\textcolor{blue}{Nano Lett.}} \textbf{24}(42), 13300–13306 (2024).

\bibitem{mintairov2025dirac}
A. M. Mintairov, V. Yu. Axenov, D. V. Lebedev, A. S. Vlasov, A. S. Frolov, E. V. Ponamarev, and V. S. Stolyarov, ``Dirac anyons in magnetophotoluminescence spectra of Wigner quantum dots: Molecular versus puddle states,''
\href{https://link.aps.org/doi/10.1103/PhysRevB.111.045410}
{\textcolor{blue}{Phys. Rev. B}} \textbf{111}(4), 045410 (2025).

\bibitem{shornikova2020magneto}
E. V. Shornikova, D. R. Yakovlev, D. O. Tolmachev, V. Yu. Ivanov, I. V. Kalitukha, V. F. Sapega, D. Kudlacik, Y. G. Kusrayev, A. A. Golovatenko, S. Shendre, et al., ``Magneto-optics of excitons interacting with magnetic ions in CdSe/CdMnS colloidal nanoplatelets,''
\href{https://pubs.acs.org/doi/full/10.1021/acsnano.0c04048}
{\textcolor{blue}{ACS Nano}} \textbf{14}(7), 9032–9041 (2020).

 \bibitem{liu2020landau}
E. Liu, J. van Baren, T. Taniguchi, K. Watanabe, Y.-C. Chang, and C. H. Lui, ``Landau-Quantized Excitonic Absorption and Luminescence in a Monolayer Valley Semiconductor,''
\href{https://link.aps.org/doi/10.1103/PhysRevLett.124.097401}
{\textcolor{blue}{Phys. Rev. Lett.}} \textbf{124}(9), 097401 (2020).

\bibitem{robert2020measurement}
C. Robert, B. Han, P. Kapuscinski, A. Delhomme, C.Faugeras, T. Amand, M.R. Molas, M. Bartos, K. Watanabe, T. Taniguchi, et al., ``Measurement of the spin-forbidden dark excitons in MoS$_2$ and MoSe$_2$ monolayers,''
\href{https://www.nature.com/articles/s41467-020-17608-4}
{\textcolor{blue}{Nat. Commun.}} \textbf{11}(1), 4037 (2020).

\bibitem{vantuan2022sixbody}
D. Van Tuan, S.-F. Shi, X. Xu, S. A. Crooker, and H. Dery, ``Six-Body and Eight-Body Exciton States in Monolayer ${\mathrm{WSe}}_{2}$,''
\href{https://link.aps.org/doi/10.1103/PhysRevLett.129.076801}
{\textcolor{blue}{Phys. Rev. Lett.}} \textbf{129}(7), 076801 (2022).

\bibitem{Li2019_DarkExciton}
Z. Li, T. Wang, C. Jin, Z. Lu, Z. Lian, Y. Meng, M. Blei, S. Gao, T. Taniguchi, K. Watanabe, \textit{et al.}, ``Emerging photoluminescence from the dark-exciton phonon replica in monolayer WSe$_2$,'' 
\href{https://www.nature.com/articles/s41467-019-10477-6}
{\textcolor{blue}{Nat. Commun.}} \textbf{10}(1), 2469 (2019).

\bibitem{Cong2015_Superfluorescence}
K. Cong, Y. Wang, J.-H. Kim, G. T. Noe, S. A. McGill, A. Belyanin, and J. Kono, ``Superfluorescence from photoexcited semiconductor quantum wells: Magnetic field, temperature, and excitation power dependence,'' 
\href{https://link.aps.org/doi/10.1103/PhysRevB.91.235448}
{\textcolor{blue}{Phys. Rev. B}} \textbf{91}(23), 235448 (2015).

\bibitem{Bryja2004_MagneticExcitons}
L. Bryja, M. Kubisa, K. Ryczko, J. Misiewicz, R. Stpniewski, M. Byszewski, M. Potemski, D. Reuter, and A. Wieck, ``Magnetic-field-induced excitons in photoluminescence from heavily doped p-type Ga$_{1-x}$Al$_x$As/GaAs single heterojunction,'' 
\href{https://www.sciencedirect.com/science/article/pii/S0921452604001206}
{\textcolor{blue}{Physica B}} \textbf{346}, 442--445 (2004).

\bibitem{zhang202426}
X. Zhang, S. Hu, L. Guo, W. Hong, Z. Wang, H. Ma, S. Zhang, J. Qin, C. Zhou, P. Gao, \textit{et al.}, ``26.86-tesla direct-current magnetic field generated with an all-REBCO superconducting magnet,'' 
\href{https://iopscience.iop.org/article/10.1088/1361-6668/ad54f9}
{\textcolor{blue}{Supercond. Sci. Technol.}} \textbf{37}(8), 085003 (2024).

\bibitem{twin2007present}
A. Twin, J. Brown, F. Domptail, R. Bateman, R. Harrison, M. Lakrimi, Z. Melhem, P. Noonan, M. Field, S. Hong, \textit{et al.}, ``Present and future applications for advanced superconducting materials in high field magnets,'' 
\href{https://ieeexplore.ieee.org/abstract/document/4277775}
{\textcolor{blue}{IEEE Trans. Appl. Supercond.}} \textbf{17}(2), 2295--2298 (2007).

\bibitem{Hollis}
W. Ma, R. Viznichenko, A. Twin, A. Varney, N. Clarke, D. Warren, R. Wotherspoon, and T. Hollis, ``A new member of high field large bore superconducting research magnets family,'' 
\href{https://iopscience.iop.org/article/10.1088/1757-899X/502/1/012104}
{\textcolor{blue}{IOP Conf. Ser.: Mater. Sci. Eng.}} \textbf{502}(1), 012104 (2019).

\bibitem{chen2012resistive}
J. Chen and M. D. Bird, ``Resistive solenoid development at the NHMFL based on irregular stacking method,'' 
\href{https://ieeexplore.ieee.org/abstract/document/6187702}
{\textcolor{blue}{IEEE Trans. Appl. Supercond.}} \textbf{22}(3), 4301704 (2012).

\bibitem{zherlitsyn2010design}
S. Zherlitsyn, T. Herrmannsdörfer, B. Wustmann, and J. Wosnitza, ``Design and performance of non-destructive pulsed magnets at the Dresden High Magnetic Field Laboratory,'' 
\href{https://ieeexplore.ieee.org/abstract/document/5451068}
{\textcolor{blue}{IEEE Trans. Appl. Supercond.}} \textbf{20}(3), 672--675 (2010).

\bibitem{murthy2007construction}
O. V. S. N. Murthy and V. Venkataraman, ``Construction and calibration of a 12 T pulsed magnet integrated with a 4 K closed-cycle refrigerator,''
\href{https://pubs.aip.org/aip/rsi/article/78/11/113905/309551}
{\textcolor{blue}{Rev. Sci. Instrum.}} \textbf{78}(11), 113905 (2007).

\bibitem{han2017pulsed}
X. Han, T. Peng, H. Ding, T. Ding, Z. Zhu, Z. Xia, J. Wang, J. Han, Z. Ouyang, Z. Wang, et al., ``The pulsed high magnetic field facility and scientific research at Wuhan National High Magnetic Field Center,''
\href{https://pubs.aip.org/aip/mre/article/2/6/278/252834/The-pulsed-high-magnetic-field-facility-and}
{\textcolor{blue}{Matter Radiat. Extremes}} \textbf{2}(6), 278--286 (2017).

\bibitem{boebinger2001national}
G. S. Boebinger, A. H. Lacerda, H. J. Schneider-Muntau, and N. Sullivan, ``The National High Magnetic Field Laboratory's pulsed magnetic field facility in Los Alamos,''
\href{https://www.sciencedirect.com/science/article/pii/S0921452600007122?via%3Dihub}
{\textcolor{blue}{Physica B: Condens. Matter}} \textbf{294}, 512--518 (2001).

\bibitem{tay2022magneto}
F. Tay, A. Baydin, F. Katsutani, and J. Kono, ``Magneto-optical spectroscopy with RAMBO: A table-top 30 T magnet,''
\href{https://journals.jps.jp/doi/10.7566/JPSJ.91.101006}
{\textcolor{blue}{J. Phys. Soc. Jpn.}} \textbf{91}(10), 101006 (2022).

\bibitem{lncmp2006toulouse}
LNCMP Team and others, ``The Toulouse pulsed magnet facility,''
\href{https://iopscience.iop.org/article/10.1088/1742-6596/51/1/146}
{\textcolor{blue}{J. Phys.: Conf. Ser.}} \textbf{51}(1), 146 (2006).

\bibitem{noe2013tabletop}
G. T. Noe II, H. Nojiri, J. Lee, G. L. Woods, J. Léotin, and J. Kono, ``A table-top, repetitive pulsed magnet for nonlinear and ultrafast spectroscopy in high magnetic fields up to 30 T,''
\href{https://doi.org/10.1063/1.4850675}
{\textcolor{blue}{Rev. Sci. Instrum.}} \textbf{84}(12), 123906 (2013).

\bibitem{naumov2010closed}
P. G. Naumov, I. S. Lyubutin, K. V. Frolov, and E. I. Demikhov, ``A closed-cycle cryostat for optical and M\"ossbauer spectroscopy in the temperature range 4.2–300 K,''
\href{https://link.springer.com/article/10.1134/S0020441210050301}
{\textcolor{blue}{Instrum. Exp. Tech.}} \textbf{53}(5), 770–776 (2010).

\bibitem{berryhill2008novel}
A. B. Berryhill, D. M. Coffey, R. W. McGhee, and E. E. Burkhardt, ``Novel integration of a 6 T cryogen-free magneto-optical system with a variable-temperature sample using a single cryocooler,''
\href{https://pubs.aip.org/aip/acp/article/985/1/1523/990605/NOVEL-INTEGRATION-OF-A-6T-CRYOGEN-FREE-MAGNETO}
{\textcolor{blue}{AIP Conf. Proc.}} \textbf{985}(1), 1523–1528 (2008).

\bibitem{goto2011optical}
A. Goto, S. Ohki, K. Hashi, and T. Shimizu, ``Optical-pumping double-nuclear-magnetic-resonance system with a Gifford--McMahon cryocooler,''
\href{https://iopscience.iop.org/article/10.1143/JJAP.50.126701}
{\textcolor{blue}{Jpn. J. Appl. Phys.}} \textbf{50}(12R), 126701 (2011).

\bibitem{low2021scanning}
D. Low, G. M. Ferguson, A. Jarjour, B. T. Schaefer, M. D. Bachmann, P. J. W. Moll, and K. C. Nowack, ``Scanning SQUID microscopy in a cryogen-free dilution refrigerator,''
\href{https://pubs.aip.org/aip/rsi/article/92/8/083704/1031323}
{\textcolor{blue}{Rev. Sci. Instrum.}} \textbf{92}(8), 083704 (2021).

\bibitem{Mukhuti-Bansal}
Kingshuk Mukhuti and Bhavtosh Bansal, ``Diamagnetism-Induced Suppression of the Resonance Energy,'' \href{https://doi.org/10.1021/acs.jpcc.3c02639}{\textcolor{blue}{
J. Phys. Chem. C} {\bf 127}, 13130 (2023)}.

\bibitem{Kratz2002_C2}
R. Kratz and P. Wyder, \textit{Pulsed Field Facilities}, in \textit{Principles of Pulsed Magnet Design},
\href{https://link.springer.com/chapter/10.1007/978-3-662-04969-3_4}
{\textcolor{blue}{Springer}} (2002).

\bibitem{Herrmannsdoerfer2003_C2}
T. Herrmannsd{\"o}rfer, H. Krug, F. Pobell, S. Zherlitsyn, H. Eschrig, J. Freudenberger, K. H. M{\"u}ller, and L. Schultz, ``The high field project at Dresden/Rossendorf: A pulsed 100 T/10 ms laboratory at an infrared free-electron-laser facility,'' 
\href{https://link.springer.com/article/10.1023/A:1025680916320}
{\textcolor{blue}{J. Low Temp. Phys.}} \textbf{133}(1), 41--59 (2003).

\bibitem{Singleton2004_C2}
J. Singleton, C. H. Mielke, A. Migliori, G. S. Boebinger, and A. H. Lacerda, ``The national high magnetic field laboratory pulsed-field facility at Los Alamos National Laboratory,'' 
\href{https://www.sciencedirect.com/science/article/pii/S0921452604001097}
{\textcolor{blue}{Physica B}} \textbf{346}, 614--617 (2004).

\bibitem{Ortenberg2001_C2}
M. von Ortenberg, N. Puhlmann, I. Stolpe, H.-U. Mueller, A. Kirste, and O. Portugall, ``The Humboldt high magnetic field center at Berlin,'' 
\href{https://www.sciencedirect.com/science/article/pii/S0921452600007225}
{\textcolor{blue}{Physica B}} \textbf{294}, 568--573 (2001).

\bibitem{Miura2001_C2}
N. Miura, Y. H. Matsuda, K. Uchida, S. Todo, T. Goto, H. Mitamura, T. Osada, and E. Ohmichi, ``New megagauss laboratory of ISSP at Kashiwa,'' 
\href{https://www.sciencedirect.com/science/article/pii/S0921452600007213}
{\textcolor{blue}{Physica B}} \textbf{294}, 562--567 (2001).

\bibitem{Bykov2001_C2}
A. I. Bykov, M. I. Dolotenko, N. P. Kolokolchikov, V. D. Selemir, and O. M. Tatsenko, ``VNIIEF achievements on ultra-high magnetic fields generation,'' 
\href{https://www.sciencedirect.com/science/article/pii/S0921452600007237}
{\textcolor{blue}{Physica B}} \textbf{294}, 574--578 (2001).

\bibitem{peng2008efficient}
T. Peng, L. Li, J. Vanacken, and F. Herlach, ``Efficient design of advanced pulsed magnets,''
\href{https://ieeexplore.ieee.org/abstract/document/4497943}
{\textcolor{blue}{IEEE Trans. Appl. Supercond.}} \textbf{18}(2), 1509--1512 (2008).

\bibitem{herlach2006experimental}
F. Herlach, T. Peng, and J. Vanacken, ``Experimental and theoretical analysis of the heat distribution in pulsed magnets,''
\href{https://ieeexplore.ieee.org/abstract/document/1643185}
{\textcolor{blue}{IEEE Trans. Appl. Supercond.}} \textbf{16}(2), 1689--1691 (2006).

\bibitem{vanacken2001pulsed}
J. Vanacken, L. Liang, K. Rosseel, W. Boon, and F. Herlach, ``Pulsed magnet design software,''
\href{https://www.sciencedirect.com/science/article/pii/S0921452600007420}
{\textcolor{blue}{Physica B}} \textbf{294}, 674--678 (2001).

\bibitem{Richardson-Smith} R. C. Richardson and E. N. Smith, {\it Experimental Techniques in Condensed Matter Physics at Low Temperatures}, Frontiers in Physics No. 67 (Addison-Wesley Pub. Co, Redwood City, Calif, 1988)

\bibitem{Patane2012_C2}
A. Patanè and N. Balkan, \textit{Semiconductor Research: Experimental Techniques}, 
\href{https://books.google.co.in/books?id=nQMsZlv2odoC}
{\textcolor{blue}{Springer}} (2012).



\bibitem{Maes-Diamond}J. Maes, K. Iakoubovskii, M. Hayne, A. Stesmans and V. V. Moshchalkov, ``Diamond as a magnetic field calibration probe,'' \href{https://iopscience.iop.org/article/10.1088/0022-3727/37/7/024}{\textcolor{blue}{J. Phys. D: Appl. Phys.} {\bf 37}, 1102 (2004).}

\bibitem{Klingshirn2012_C5} C. F. Klingshirn, Semiconductor Optics, Springer Science \& Business Media (2012).

\bibitem{vrehen1968interband}
Q. H. F. Vrehen, 
``Interband magneto-optical absorption in gallium arsenide,''
\href{https://www.sciencedirect.com/science/article/pii/0022369768902631}
{\textcolor{blue}{J. Phys. Chem. Solids}} \textbf{29}(1), 129--141 (1968).



\bibitem{moss1961optical}
T. S. Moss, 
``Optical absorption edge in GaAs and its dependence on electric field,''
\href{https://pubs.aip.org/aip/jap/article/32/10/2136/377102/Optical-Absorption-Edge-in-GaAs-and-Its-Dependence}
{\textcolor{blue}{J. Appl. Phys.}} \textbf{32}(10), 2136--2139 (1961).

\bibitem{wang2017structural}
K.-H. Wang, L.-C. Li, M. Shellaiah, and K. W. Sun, ``Structural and photophysical properties of methylammonium lead tribromide (MAPbBr$_3$) single crystals,''
\href{https://www.nature.com/articles/s41598-017-13571-1}
{\textcolor{blue}{Sci. Rep.}} \textbf{7}(1), 13643 (2017).

\bibitem{baranowski2024polaronic}
M. Baranowski, A. Nowok, K. Galkowski, M. Dyksik, A. Surrente, D. Maude, 
M. Zacharias, G. Volonakis, S. D. Stranks, J. Even, \textit{et al.}, 
``Polaronic mass enhancement and polaronic excitons in metal halide perovskites,''
\href{https://pubs.acs.org/doi/full/10.1021/acsenergylett.4c00905}
{\textcolor{blue}{ACS Energy Lett.}} \textbf{9}(6), 2696--2702 (2024).

\bibitem{galkowski2016determination}
K. Galkowski, A. Mitioglu, A. Miyata, P. Plochocka, O. Portugall, 
G. E. Eperon, J. T.-W. Wang, T. Stergiopoulos, S. D. Stranks, 
H. J. Snaith, \textit{et al.}, 
``Determination of the exciton binding energy and effective masses for methylammonium and formamidinium lead tri-halide perovskite semiconductors,''
\href{https://pubs.rsc.org/en/content/articlehtml/2016/ee/c5ee03435c}
{\textcolor{blue}{Energy Environ. Sci.}} \textbf{9}(3), 962--970 (2016).

\bibitem{yang2017unraveling}
Z. Yang, A. Surrente, K. Galkowski, N. Bruyant, D. K. Maude, A. A. Haghighirad, H. J. Snaith, P. Plochocka, and R. J. Nicholas, ``Unraveling the exciton binding energy and the dielectric constant in single-crystal methylammonium lead triiodide perovskite,'' 
\href{https://pubs.acs.org/doi/10.1021/acs.jpclett.7b00524}
{\textcolor{blue}{J. Phys. Chem. Lett.}} \textbf{8}(8), 1851--1855 (2017).

\bibitem{tilchin2016hydrogen}
J. Tilchin, D. N. Dirin, G. I. Maikov, A. Sashchiuk, M. V. Kovalenko, and E. Lifshitz, ``Hydrogen-like Wannier--Mott excitons in single crystal of methylammonium lead bromide perovskite,'' 
\href{https://pubs.acs.org/doi/10.1021/acsnano.6b02734}
{\textcolor{blue}{ACS Nano}} \textbf{10}(6), 6363--6371 (2016).

\bibitem{PhysRevMaterials.8.034601}
I. V. Zhevstovskikh, N. S. Averkiev, M. N. Sarychev, 
O. I. Semenova, and O. E. Tereshchenko, 
``Exciton binding energy and the origin of yellow-light emission in ${\mathrm{CH}}_{3}{\mathrm{NH}}_{3}{\mathrm{PbBr}}_{3}$ single crystals,''
\href{https://link.aps.org/doi/10.1103/PhysRevMaterials.8.034601}
{\textcolor{blue}{Phys. Rev. Mater.}} \textbf{8}(3), 034601 (2024).


\end{thebibliography}
 \end{document}